\documentclass[twocolumn,english]{IEEEtran}
\usepackage[T1]{fontenc}
\usepackage[latin9]{inputenc}
\usepackage{color}
\usepackage{babel}
\usepackage{array}
\usepackage{refstyle}
\usepackage{booktabs}
\usepackage{multirow}
\usepackage{amsmath}
\usepackage{amssymb}
\usepackage{graphicx}
\usepackage{setspace}
\usepackage[unicode=true,pdfusetitle,
 bookmarks=true,bookmarksnumbered=false,bookmarksopen=false,
 breaklinks=false,pdfborder={0 0 1},backref=false,colorlinks=true]
 {hyperref}

\makeatletter

\AtBeginDocument{\providecommand\secref[1]{\ref{sec:#1}}}
\AtBeginDocument{\providecommand\subsecref[1]{\ref{subsec:#1}}}
\AtBeginDocument{\providecommand\figref[1]{\ref{fig:#1}}}
\providecommand{\tabularnewline}{\\}
\RS@ifundefined{subsecref}
  {\newref{subsec}{name = \RSsectxt}}
  {}
\RS@ifundefined{thmref}
  {\def\RSthmtxt{theorem~}\newref{thm}{name = \RSthmtxt}}
  {}
\RS@ifundefined{lemref}
  {\def\RSlemtxt{lemma~}\newref{lem}{name = \RSlemtxt}}
  {}

\date{}

\makeatother

\begin{document}

\title{Supervised Q-walk for Learning Vector Representation of Nodes in
Networks}

\author{Naimish Agarwal, IIIT-Allahabad (\textit{irm2013013@iiita.ac.in})\\
G.C. Nandi, IIIT-Allahabad (\textit{gcnandi@iiita.ac.in)}}
\maketitle
\begin{abstract}
Automatic feature learning algorithms are at the forefront of modern
day machine learning research. We present a novel algorithm\textit{,
supervised Q-walk},\textit{ }which applies Q-learning to generate
random walks on graphs such that the walks prove to be useful for
learning node features suitable for tackling with the node classification
problem.\textit{ }We present another novel algorithm, \textit{k-hops
neighborhood based confidence values learner}, which learns confidence
values of labels for unlabelled nodes in the network without first
learning the node embedding. These confidence values aid in learning
an apt reward function for Q-learning. 

We demonstrate the efficacy of supervised Q-walk approach over existing
state-of-the-art random walk based node embedding learners in solving
the single / multi-label multi-class node classification problem using
several real world datasets. 

Summarising, our approach represents a novel state-of-the-art technique
to learn features, for nodes in networks, tailor-made for dealing
with the node classification problem.
\end{abstract}


\section{Introduction}

Consider a social network of users where users have professions like
scientist, manager, student, etc. Each user can be connected to sets
of users from different professions. The problem is to predict the
profession of users whose profession information is missing based
upon their context in the network. Such a problem belongs to the class
of problems called node classification problem. 

In node classification problem, the task is to predict the labels
of those nodes in networks whose label information is missing. A node
can have one or more labels associated with it, e.g. if the nodes
are people, then the labels can be student, artist, dancer, etc. This
problem seems to be solvable using supervised machine learning techniques.
The downside is that we do not know the required set of features which
need to be feeded into the machine learning system for performing
node classification.

In our work, we have developed a technique for learning features of
nodes in networks, i.e. node embeddings. Our technique provides a
supervised adaptation to node2vec \cite{Grover:2016:NSF:2939672.2939754}
- a semi-supervised learning algorithm to learn continuous valued
features of nodes in networks. 

Our algorithm is based on a simple intuition. We want that the nodes
which have same labels should have very similar embeddings. Therefore,
we would like to perform random walks on the nodes in the networks
such that if the nodes on the random walk are $u_{1},u_{2},u_{3},...,u_{w_{l}}$
where $w_{l}$ is the walk length, then $u_{i}$ is very similar to
$u_{i-1}$ where $i=2,3,4,...,w_{l}$.

To achieve the above intuition, we have laid out a two step approach
to perform the random walks:
\begin{enumerate}
\item We associate confidence values with all node-label pairs. The confidence
values just give us a hint about the tentative labels for nodes. 
\item We first associate a reward function with every edge in the network.
These reward functions are used by the Q-learning algorithm in learning
the Q-values for each node-edge pair. These Q-values then guide the
random walks in a way such that in the ideal scenario if the random
walk is $u_{1},u_{2},u_{3},...,u_{w_{l}}$, then the actual labels
associated with the nodes are $l,l,l,...,l$.
\end{enumerate}
The generated random walks are treated as sentences in a document.
We then apply Skip-gram \cite{DBLP:journals/corr/abs-1301-3781} with
Negative Sampling \cite{NIPS2013_5021} to learn the node embeddings.
These node embeddings are then used to train a classifier for checking
the goodness of the learnt embeddings.

The related work is discussed in \secref{Related-Work}.The details
of the approach are given in \secref{Learning-Vector-Representation}.
The experimental details are mentioned in \secref{Experiments}. The
conclusion and future work are mentioned in \secref{Conclusion} and
\secref{Future-Work} respectively. 

\section{Related Work\label{sec:Related-Work}}

Graph Analytics field is pacing up due to the growth of large datasets
in social network analysis \cite{NIPS2012_4532}\cite{Backstrom:2006:GFL:1150402.1150412}\cite{DBLP:journals/corr/abs-0810-1355},
communication networks \cite{Leskovec:2007:GED:1217299.1217301}\cite{Leskovec:2010:SNS:1753326.1753532},
etc. The area of node classification \cite{DBLP:journals/corr/abs-1101-3291}
has been approached earlier from different perspectives like factorization
based approaches, random walk based approaches, etc. 

Factorization based techniques represent the edges in networks as
matrices. These matrices are factorized to obtain the embeddings.
The matrix representation and its factorization are done using various
techniques \cite{Roweis00nonlineardimensionality}\cite{Belkin01laplacianeigenmaps}\cite{Ahmed:2013:DLN:2488388.2488393}\cite{Cao:2015:GLG:2806416.2806512}\cite{Ou:2016:ATP:2939672.2939751}.
These methods may suffer from scalability issues for large graph datasets
and sparse matrix representations need special attention. 

Random walk based approaches perform random walks on networks to obtain
the embeddings. Two popular techniques are DeepWalk \cite{Perozzi:2014:DOL:2623330.2623732}
and node2vec \cite{Grover:2016:NSF:2939672.2939754}. node2vec \cite{Grover:2016:NSF:2939672.2939754}
is a semi-supervised algorithmic framework which showcases strategies
to perform random walks such that nodes which are homophilic and/or
structurally equivalent end up getting similar embeddings. The random
walks are guided by a heuristic which involves computing distance
of the next possible nodes from the previous node given the current
node. DeepWalk can be considered as a special case of node2vec with
$p=1$ and $q=1$ where $p,q$ are hyperparameters in node2vec which
decide the tradeoff between depth-first and breadth-first sampling. 

Our approach falls under random walk based approaches. We compare
our approach against node2vec in \secref{Experiments}. Instead of
using some hand-crafted random walk based approach, we decided to
learn how to do random walks using reinforcement learning. We perform
random walks such that nodes which have same labels but are not necessarily
structurally equivalent, end up getting embeddings close to one another
in the embedding space. The random walks are guided by the Q-values
of the node-action pairs.

\section{Learning Vector Representation of Nodes\label{sec:Learning-Vector-Representation}}

In \subsecref{Problem-Definition}, we define the problem formally.
In \subsecref{K-hops-Neighborhood-based-confidence-values-learner},
we propose a k-hops neighborhood based confidence values learner which
learns the confidence values of labels for unlabelled nodes in the
network. Using the learnt confidence values, we devise a reward function,
which aids Q-walk, described in \subsecref{Q-Walk}, to do random
walks. The generated random walks are then fed into word2vec, which
is briefly described in \subsecref{Word2vec}, to get the vector representation
of nodes.

\subsection{Problem Definition\label{subsec:Problem-Definition}}

Consider $G=(V,E)$ where $G$ can be any (un)directed, (un)weighted
simple graph. We ignore self-loops and parallel edges. $V$ is the
set of vertices and $E$ is the set of edges. $V=V_{labelled}\cup V_{unlabelled}$
and $V_{labelled}\cap V_{unlabelled}=\phi$ where $V_{labelled}$
is the set of labelled vertices, $V_{unlabelled}$ is the set of unlabelled
vertices and $\phi$ denotes empty set. Let ${\displaystyle v_{l}=\frac{|V_{labelled}|}{|V|}}$.

We want to learn a mapping $f:V\longrightarrow\mathbb{{R}}^{d}$ such
that for any $u,v\in V$ if $u$ and $v$ have same labels, then $||f(u)-f(v)||_{2}$
distance is minimum. We use the same objective function and assumtions
- conditional independence and symmetry in feature space, as used
in node2vec \cite{Grover:2016:NSF:2939672.2939754}.

\begin{equation}
O_{f}=\max_{f}\,{\displaystyle \sum_{u\in V}log\,Pr(N_{S}(u)|f(u))}\label{eq:objective}
\end{equation}

In (\ref{eq:objective}), we are maximizing over the log-probability
of observing a network neighborhood $N_{S}(u)$ obtained by the sampling
strategy $S$, which, in our case, is described in \subsecref{Q-Walk},
starting at node $u$. In \cite{Grover:2016:NSF:2939672.2939754},
the authors have derived that 
\begin{equation}
O_{f}\propto\sum_{u\in V}{\displaystyle \sum_{n_{i}\in N_{S}(u)}f(n_{i}).f(u)}
\end{equation}

The use of (\ref{eq:objective}) is justified in our case since we
want to minimize $||f(u)-f(v)||_{2}$ which is equivalent to maximizing
$f(u).f(v)$.

\subsection{K-hops neighborhood based confidence values learner \label{subsec:K-hops-Neighborhood-based-confidence-values-learner}}

This algorithm is motivated by homophily \cite{mcpherson2001birds}
in networks. Entitites of similar kind tend to stay together, e.g.
friends who share same interest, people in the same profession, etc.
We use this heuristic to find the confidence values of labels for
unlabelled nodes in the graph. It is imperative to compute the confidence
values, since in their absence, the agent, as used in \subsecref{Q-Walk},
would get confused in chosing the appropriate direction of the random
walk.
\begin{equation}
N(u,k)=\{x\,|\,x\in V\,\textnormal{and}\,x\,\textnormal{is}\,k\textnormal{-}\textnormal{\textnormal{hops away from}}\,u\}\label{eq:formal-neighborhood}
\end{equation}

In (\ref{eq:formal-neighborhood}), $N(u,k)$ is the k-hops neighborhood
for $u\in V$. For directed $G$, $k$-hops are based on the outgoing
edges from $u$.

\begin{equation}
L=\{l_{i}\,|\,l_{i}\,\textnormal{is\,a\,label\,for}\,i=1,2,...,|L|\}\label{eq:formal-labels}
\end{equation}

In (\ref{eq:formal-labels}), $L$ is the set of all labels in $G$
and $|L|$ is the cardinality of $L$. 

\begin{equation}
L_{actual}(u)=\{l\,|\,u\,\textnormal{is labelled}\,l\,\textnormal{where}\,l\in L\}\label{eq:formal-labels-actual}
\end{equation}

In (\ref{eq:formal-labels-actual}), $L_{actual}(u)$ is the set of
labels associated with $u\in V_{labelled}$.

\begin{equation}
t=0,1,2,...,T\label{eq:confidence-iterator}
\end{equation}

In (\ref{eq:confidence-iterator}), $t$ is the iteration counter
for computing the confidence values and $T$ is the maximum number
of such iterations.

\begin{equation}
C_{0}(u,l)=\begin{cases}
1 & u\in V_{labelled}\,\textnormal{and}\,l\in L_{actual}(u)\\
0 & u\in V_{labelled}\,\textnormal{and}\,l\notin L_{actual}(u)\\
0 & u\in V_{unlabelled}\,\forall l\in L
\end{cases}\label{eq:initial-confidence-values}
\end{equation}

In (\ref{eq:initial-confidence-values}), $C_{0}(u,l)$ is the initial
confidence value associated with $u\in V$ computed $\forall l\in L$.

\begin{multline}
C_{t}(u,l)=\begin{cases}
C_{t-1}(u,l) & u\in V_{labelled}\\
\frac{{\displaystyle \sum_{i}^{|N(u,k)|}}C_{t-1}(x_{i},l)}{|N(u,k)|} & u\in V_{unlabelled}
\end{cases}\label{eq:t-confidence-values}
\end{multline}

In (\ref{eq:t-confidence-values}), $C_{t}(u,l)$ is the confidence
value of $u\in V$ for label $l\in L$ at iteration $t$ and $x_{i}\in N(u,k)$. 

$C_{T}(u,l)$ is the confidence with which we can state that $u\in V$
has label $l\in L$. We denote $k$ in $k$-hops neighborhood as $k_{HN}$. 

\subsection{Supervised Q-walk \label{subsec:Q-Walk}}

To generate random walks, we look at the graph as a Markov Decision
Process (MDP) \cite{citeulike:9256539}, where each node $u\in V$
is a state, outgoing edges from $u$ are the actions possible at $u$.
The probability of reaching the neighbor $u'$ by taking the action
$(u,u')$ is $1$. Imagine an agent at $u$ which has to decide the
next node $u'$. To aid the agent in taking the right decision, we
perform Q-learning \cite{Watkins1992}. So, the agent decides $u'$
based upon the Q-value $Q(u,(u,u'))$. Hence, the generated random
walks are called Q-walks. The Q-walks are supervised because the reward
function (\ref{eq:reward}) depends on the confidence values, learnt
in a supervised way, as per \subsecref{K-hops-Neighborhood-based-confidence-values-learner}.

\begin{equation}
R(u,a,u')=-{\displaystyle \sum_{i=1}^{|L|}}|C_{T}(u',l_{i})-C_{T}(u,l_{i})|\label{eq:reward}
\end{equation}

In (\ref{eq:reward}), $R(u,a,u')$ is the reward obtained when the
agent moves from $u$ to $u'$ along the edge $a$ where $u,\,u'\in V\,,\,a=(u,u')\in E\,\textnormal{and}\,l_{i}\in L$.
The closer the reward is to $0$, the more similar $u'$ is to $u$.

\begin{equation}
j=0,1,2,...,J\label{eq:iteration-q-walk}
\end{equation}

In (\ref{eq:iteration-q-walk}), $j$ is the iteration counter in
Q-learning and $J$ is the maximum number of such iterations.

\begin{equation}
Q_{0}(u,a)=0\,\forall u\in V\,\textnormal{and}\,a\in E\label{eq:initial-q-values}
\end{equation}

In (\ref{eq:initial-q-values}), $Q_{0}(u,a)$ is the initial Q-value
for the node-action pair $(u,a)$.

\begin{equation}
\alpha_{j}=\frac{\alpha_{j-1}}{1+j}\label{eq:alpha-update}
\end{equation}

\begin{multline}
Q_{j}(u,a)=Q_{j-1}(u,a)+\alpha_{j}(R(u,a,u')+\\
\gamma\max_{a'}Q_{j-1}(u',a')-Q_{j-1}(u,a))\label{eq:q_learning}
\end{multline}

In (\ref{eq:alpha-update}) and (\ref{eq:q_learning}), $\alpha_{j}$
is the learning rate for epoch $j$, $\alpha_{0}$ is user-defined,
and $\gamma$ is the discount factor. In (\ref{eq:alpha-update}),
we are updating the learning rate $\alpha$ at each iteration such
that its value decreases with number of iterations and the difference
between $Q_{j}(u,a)$ and $Q_{j-1}(u,a)$ diminishes over time, in
other words, Q-values converge.

\begin{equation}
r_{n}\sim\textnormal{Uniform}(0,1)\label{eq:get_random_number}
\end{equation}

\begin{equation}
a=\begin{cases}
\underset{a'}{\arg\max\,}Q_{J}(u',a') & r_{n}\leq p_{Q}\\
(u,\textnormal{random}(u'))\in E & r_{n}>p_{Q}
\end{cases}\label{eq:get_action}
\end{equation}

The agent needs to decide the action when it is at $u$. It first
samples a random number as per (\ref{eq:get_random_number}), then
it decides upon the action as per (\ref{eq:get_action}). In (\ref{eq:get_action}),
$p_{Q}$ is the exploitation probability and $u'\in N(u,1)$. We generate
$r$ random walks of length $w_{l}$ each $\forall u\in V$.

\subsection{word2vec\label{subsec:Word2vec}}

The generated random walks can be considered as sentences in a text
document. As per (\ref{eq:objective}), we have to maximize the probability
of the context nodes given the node $u\in V$. To achieve this, we
use Skip-gram \cite{DBLP:journals/corr/abs-1301-3781} with Negative
Sampling \cite{NIPS2013_5021}. We denote the context window size
as $w_{s}$ and number of epochs as $e$. 

\section{Experiments\label{sec:Experiments}}

\begin{figure}
\includegraphics[width=1\linewidth]{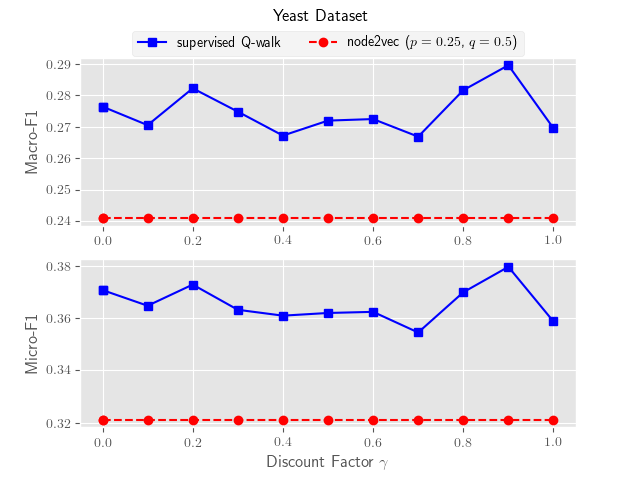}

\caption{Mean F1 scores of supervised Q-walk against the baseline node2vec
for different settings of the discount factor $\gamma$ with fixed
$p_{Q}=0.8$, $k_{HN}=1$, $r=24$, $w_{l}=100$, $d=256$, $w_{s}=20$,
$e=100$, $v_{l}=0.8$, $\alpha_{0}=1$, $J=100$, $T=3$, $k_{NN}=3$.\label{fig:discount-factor}}
\end{figure}

\begin{figure}

\includegraphics[width=1\linewidth]{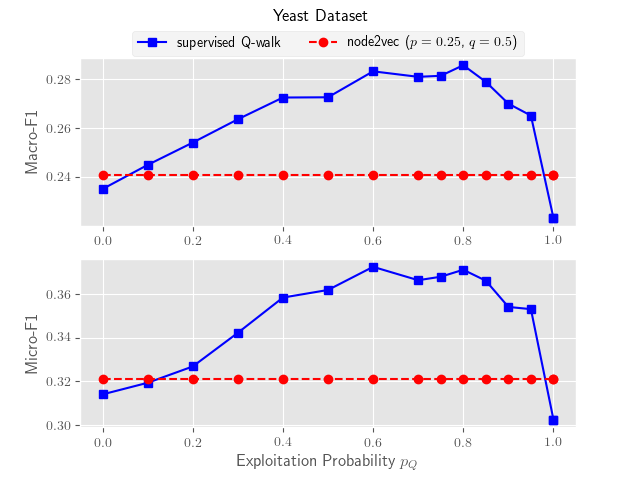}

\caption{Mean F1 scores of supervised Q-walk against baseline node2vec for
different settings of the exploitation probability $p_{Q}$ with fixed
$k_{HN}=1$, $r=24$, $w_{l}=100$, $d=256$, $w_{s}=20$, $e=100$,
$v_{l}=0.8$, $\alpha_{0}=1$, $J=100$, $T=3$, $k_{NN}=3$, $\gamma=0.9$.\label{fig:exploitation-probability}}
\end{figure}

\begin{figure}
\includegraphics[width=1\linewidth]{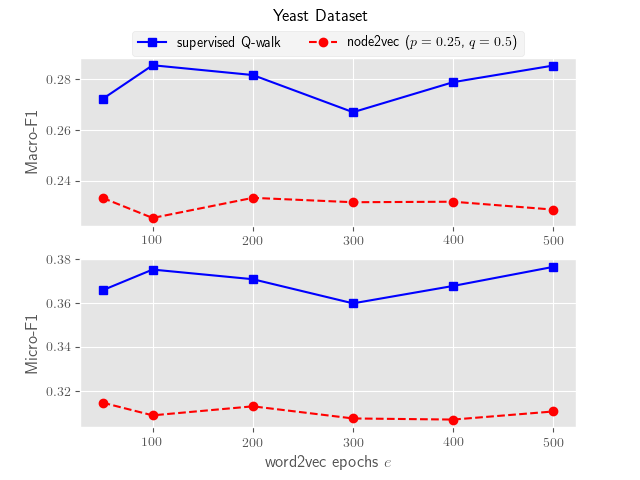}\caption{Mean F1 scores of supervised Q-walk against baseline node2vec for
different settings of the word2vec epochs $e$ with fixed $k_{HN}=1$,
$r=24$, $w_{l}=100$, $d=256$, $w_{s}=20$, $v_{l}=0.8$, $\alpha_{0}=1$,
$J=100$, $T=3$, $k_{NN}=3$, $\gamma=0.9$, $p_{Q}=0.8$. \label{fig:word2vec-epochs}}

\end{figure}

\begin{figure}

\includegraphics[width=1\linewidth]{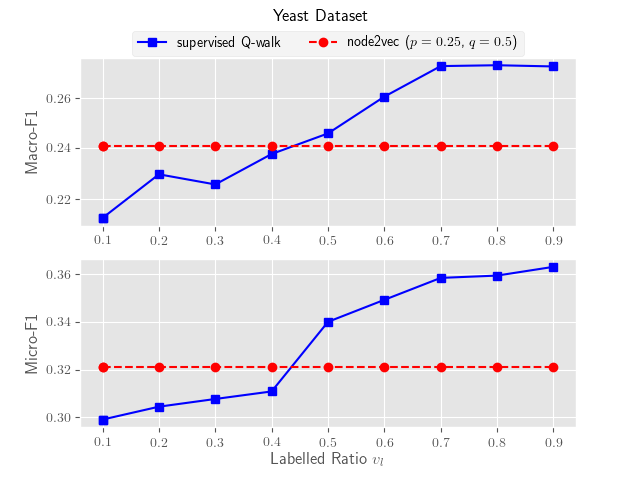}\caption{Mean F1 scores of supervised Q-walk against baseline node2vec for
different settings of the ratio of labelled nodes $v_{l}$, used for
learning confidence values as per \subsecref{K-hops-Neighborhood-based-confidence-values-learner}
with fixed $k_{HN}=1$, $r=24$, $w_{l}=100$, $d=256$, $\alpha_{0}=1$,
$J=100$, $T=3$, $k_{NN}=3$, $w_{s}=20,\gamma=0.9$, $p_{Q}=0.8$,
$e=100$.\label{fig:labelled-ratio}}

\end{figure}

\begin{figure}
\includegraphics[width=1\linewidth]{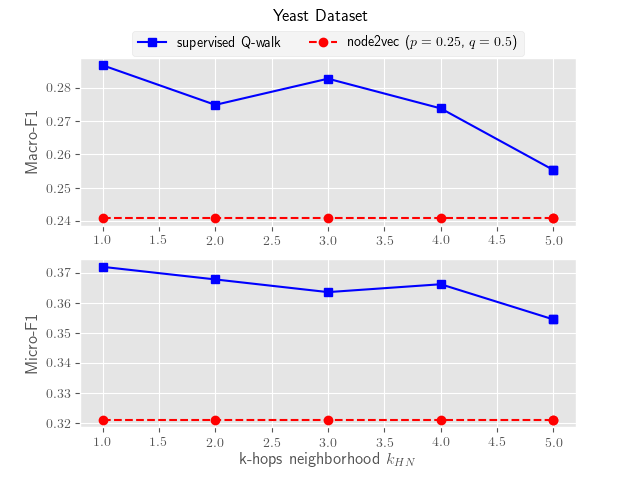}\caption{Mean F1 scores of supervised Q-walk against baseline node2vec for
different settings of $k_{HN}$, used for learning confidence values
as per \subsecref{K-hops-Neighborhood-based-confidence-values-learner}
with fixed $r=24$, $w_{l}=100$, $d=256$, $\alpha_{0}=1$, $J=100$,
$T=3$, $k_{NN}=3$, $\gamma=0.9$, $p_{Q}=0.8$, $e=100$, $w_{s}=20,v_{l}=0.8$.\label{fig:k-hops}}

\end{figure}

\begin{figure}

\includegraphics[width=1\linewidth]{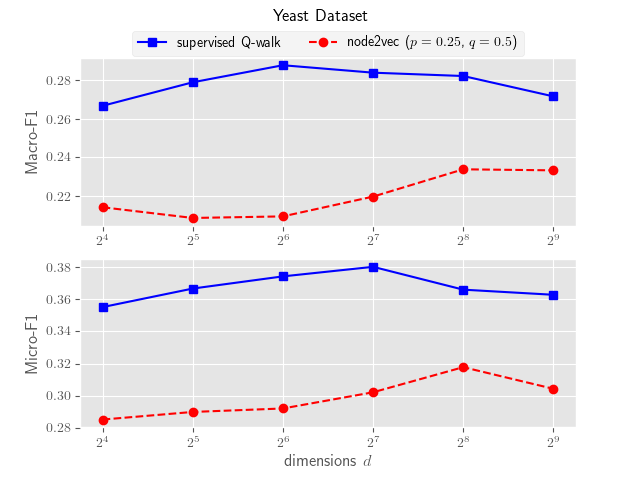}\caption{Mean F1 scores of supervised Q-walk against baseline node2vec for
different settings of node features representation dimensions $d$
with fixed $r=24$, $w_{l}=100$,$\alpha_{0}=1$, $J=100$, $T=3$,
$k_{NN}=3$, $\gamma=0.9$, $p_{Q}=0.8$, $e=100$, $v_{l}=0.8$,
$k_{HN}=1,w_{s}=20$. \label{fig:dimensions}}

\end{figure}

\begin{figure}
\includegraphics[width=1\linewidth]{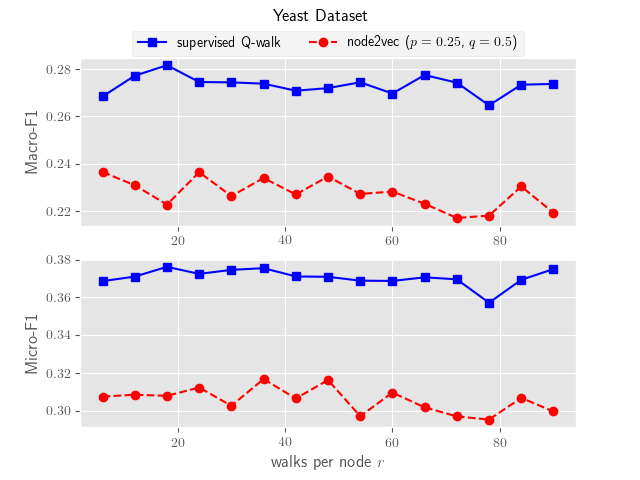}\caption{Mean F1 scores of supervised Q-walk against baseline node2vec for
different settings of number of walks per node $r$ with fixed $w_{l}=100$,$\alpha_{0}=1$,
$J=100$, $T=3$, $k_{NN}=3$, $\gamma=0.9$, $p_{Q}=0.8$, $e=100$,
$v_{l}=0.8$, $k_{HN}=1$, $d=256,w_{s}=20$.\label{fig:walks-per-node}}

\end{figure}

\begin{figure}
\includegraphics[width=1\linewidth]{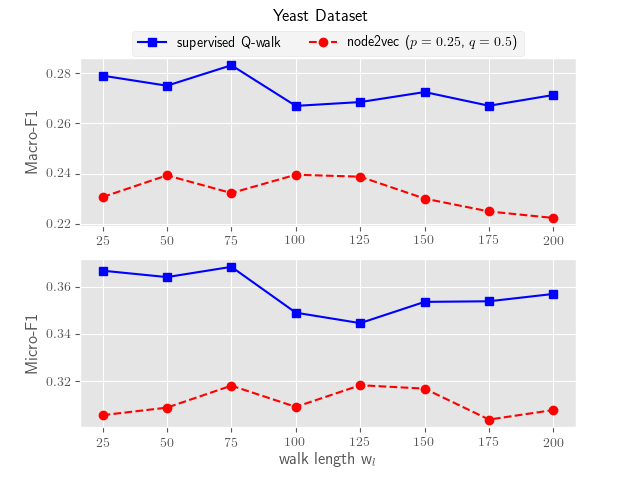}

\caption{Mean F1 scores of supervised Q-walk against baseline node2vec for
different settings of walk length $w_{l}$ with fixed $r=24,\alpha_{0}=1$,
$J=100$, $T=3$, $k_{NN}=3$, $\gamma=0.9$, $p_{Q}=0.8$, $e=100$,
$v_{l}=0.8$, $k_{HN}=1$, $d=256,w_{s}=20$. \label{fig:walk-length}}
\end{figure}

\begin{figure}

\includegraphics[width=1\linewidth]{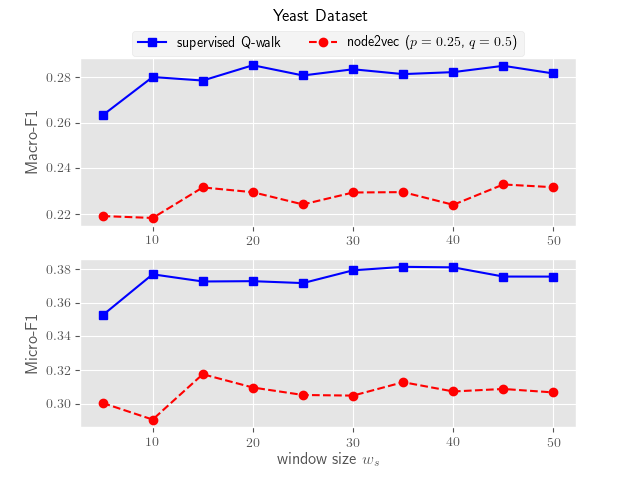}

\caption{Mean F1 scores of supervised Q-walk against baseline node2vec for
different settings of window size $w_{s}$ with fixed $r=24,\alpha_{0}=1$,
$J=100$, $T=3$, $k_{NN}=3$, $\gamma=0.9$, $p_{Q}=0.8$, $e=100$,
$v_{l}=0.8$, $k_{HN}=1$, $d=256,w_{l}=24$. \label{fig:window-size}}

\end{figure}

\begin{figure}
\includegraphics[width=1\linewidth]{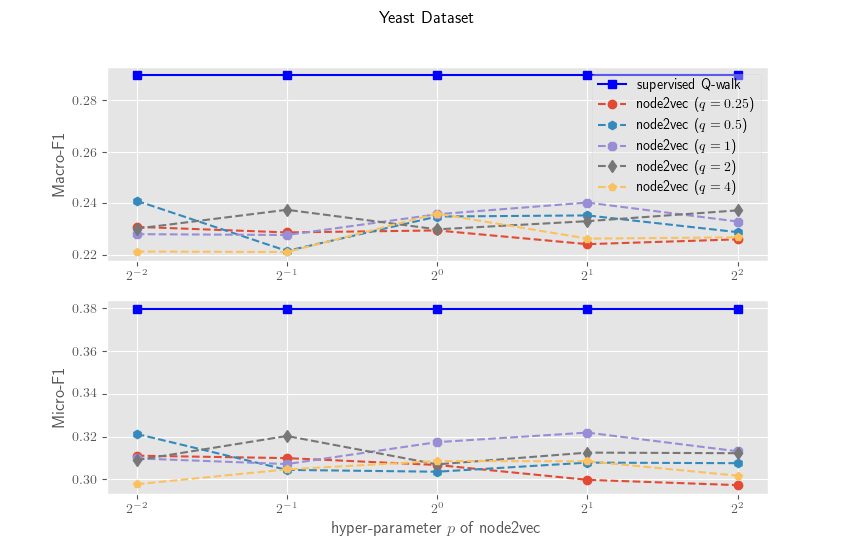}

\caption{Mean F1 scores of node2vec against supervised Q-walk for different
values of $p$ and $q$ with fixed $p_{Q}=0.8$, $k_{HN}=1$, $r=24$,
$w_{l}=100$, $d=256$, $w_{s}=20$, $e=100$, $v_{l}=0.8$, $\alpha_{0}=1$,
$J=100$, $T=3$, $k_{NN}=3$, $\gamma=0.9,w_{s}=20$.\label{fig:node2vec-p}}
\end{figure}

\begin{table*}
\caption{Mean F1 scores of supervised Q-walk and node2vec on different datasets
with fixed $p_{Q}=0.8$, $k_{HN}=1$, $r=24$, $w_{l}=100$, $d=256$,
$w_{s}=20$, $e=100$, $v_{l}=0.8$, $\alpha_{0}=1$, $J=100$, $T=3$,
$k_{NN}=3$, $\gamma=0.9$, $p=0.25$, $q=0.5$.\label{tab:f1 scores}}
\begin{singlespace}
\centering{}%
\begin{tabular*}{1\linewidth}{@{\extracolsep{\fill}}ccccccc}
\cmidrule{2-7} 
 & \multicolumn{2}{c}{Supervised Q-walk} & \multicolumn{2}{c}{node2vec } & \multicolumn{2}{c}{\% improvement}\tabularnewline
\cmidrule{2-7} 
 & Macro-F1 & Micro-F1 & Macro-F1 & Micro-F1 & Macro-F1 & Micro-F1\tabularnewline
\midrule
\multirow{1}{*}{Yeast} & \textbf{0.2896} & \textbf{0.3797} & 0.2409 & 0.3212 & \textbf{20.21\%} & \textbf{18.21\%}\tabularnewline
\midrule 
\multirow{1}{*}{BlogCatalog} & \textbf{0.4051} & \textbf{0.5420} & 0.1984 & 0.3062 & \textbf{104.18\%} & \textbf{77.01\%}\tabularnewline
\midrule 
\multirow{1}{*}{Flickr} & \textbf{0.4340} & \textbf{0.5505} & 0.2087 & 0.2853 & \textbf{107.95\%} & \textbf{92.95\%}\tabularnewline
\bottomrule
\end{tabular*}
\end{singlespace}
\end{table*}

\subsection{Datasets}

We provide a brief overview of the datasets which were used to perform
the experiments.

\subsubsection{Yeast}

Yeast \cite{citeulike:1007064} dataset is a protein-protein interaction
network of budding yeast. It has $2361$ nodes, $7182$ edges and
$13$ classes. It is an undirected and unweighted network. It offers
a single-label multi-class classification problem.

\subsubsection{BlogCatalog}

BlogCatalog \cite{Tang:2009:RLV:1557019.1557109} dataset is a social
network of bloggers. The labels associated with the bloggers refer
to their interests which are obtained from the metadata information
as available on BlogCatalog site. It consists of $10312$ nodes, $333983$
edges and $39$ classes. It is an undirected and unweighted network.
It offers a multi-label multi-class classification problem.

\subsubsection{Flickr}

Flickr \cite{Tang:2009:RLV:1557019.1557109} dataset is a contact
network of users on Flickr site. The labels associated with the users
refer to their groups. It consists of $80513$ nodes, $5899882$ edges
and $195$ classes. It is an undirected and unweighted network. It
offers a multi-label multi-class classification problem.

\subsection{Performance Evaluation}

We evaluate the performance by first computing the vector representation
of nodes in the network using both node2vec and supervised Q-walk
for specific hyperparameter settings. Then, we compute the mean of
the macro and micro F1 scores obtained by performing 5 fold cross
validation using k-nearest neighbors (k-NN) \cite{Guo2003} classifier.
We use k-NN for a couple of reasons. First, we are interested in showcasing
that our learnt embeddings are similar for nodes with same labels,
and such similarity can be measured by finding euclidean distance
between the node embeddings of the concerned nodes, and this is in
accordance with \subsecref{Problem-Definition}. Second, it is a non-linear
classifier, therefore, the learnt embeddings need not be linearly
separable for getting better classification performance. We denote
$k$ in k-NN as $k_{NN}$. 

\subsection{Results}

It can be observed in \figref{discount-factor} that values of $\gamma$
around $0.9$ provide higher performance than other values of $\gamma$
on the Yeast dataset. For $\gamma=0$, the F1 scores are better than
a number of other $\gamma$ values which signifies the aptness of
the reward function (\ref{eq:reward}), since (\ref{eq:q_learning})
is then modified to $Q_{j}(u,a)=(1-\alpha_{j})Q_{j-1}(u,a)+\alpha_{j}R(u,a,u')$,
which does not include $Q_{j-1}(u',a')$. For $\gamma=0$, on BlogCatalog
dataset, supervised Q-walk approach achieves mean Macro-F1 score of
$0.4198$ and mean Micro-F1 score of $0.5528$ which are higher than
the F1 scores for $\gamma=0.9$ as shown in table \ref{tab:f1 scores}.

In \figref{exploitation-probability}, we can observe the tradeoff
involved between exploration and exploitation. $p_{Q}=0$ means that
we do not make use of the learnt Q-values and have instead resorted
to randomly explore the network, thereby, leading to poor performance.
$p_{Q}=1$ means that we always make the greedy choice by opting for
the action which yields maximum Q-value. This is complete exploitation
policy which again leads to poor performance. $p_{Q}=0.8$ rightly
balances the tradeoff between exploration and exploitation, thereby,
leading to higher F1 scores compared with other values of $p_{Q}$.

In \figref{word2vec-epochs}, we can observe that supervised Q-walk
takes small number of epochs to give high F1 scores. In the case of
Yeast dataset, $e=100$ was sufficient. There exists randomness in
the curves which can be attributed to different weight initializations
in word2vec in different runs of the experiment. The F1 scores stop
improving after $e=100$. It signifies that the model is trained and
starts overfitting for $e>100$.

In \figref{labelled-ratio}, we can observe that supervised Q-walk
approach performs better than node2vec given we use around and above
50\% labelled nodes for learning the confidence values as per \subsecref{K-hops-Neighborhood-based-confidence-values-learner}.
For lower values of $v_{l}$, the performance is poor because the
k-hops neighborhood based confidence values learner did not get enough
labelled nodes for properly learning the confidence values. For higher
values of $v_{l}$, we can see that the performance improves. In a
real world setting, it may not be necessary that we get $v_{l}=0.9$
to train upon, instead, for all other experiments we choose $v_{l}=0.8$
since $80\%-20\%$ split of training and test data is generally used
in machine learning.

It can be observed in \figref{k-hops} that $k_{HN}=1$ is sufficient
for learning good confidence values, in other words, the confidence
values can be determined by just looking at the immediate neighbors
of any node $u\in V$.

It can be observed in \figref{dimensions} that $d\in[2^{5},2^{8}]$
give high F1 scores. Supervised Q-walk approach gives better F1 scores
than node2vec for different settings of dimensions.

It can be observed in \figref{walks-per-node} that supervised Q-walk
approach gives high F1 scores for $r=18$ while node2vec gives high
F1 scores for $r=24$ and low for $r=18$. So, our approach takes
lesser number of walks per node to give a better performance than
node2vec. 

It can be observed in \figref{walk-length} and \figref{window-size}
that supervised Q-walk approach gives better F1 scores than node2vec.
Values of $w_{l}$ around $75$ and values of $w_{s}$ around $20$
are good enough for our approach on the Yeast dataset.

It can be observed in \figref{node2vec-p} that supervised Q-walk
approach performs better than node2vec hyperparameterised by different
combinations of $p$ and $q$ which control the degree of Depth First
Sampling (DFS) and Breadth First Sampling (BFS).

The mean F1 scores are calculated for specific hyperparameter settings
which leaves room for improvement by fine tuning the hyperparameters
through cross validation with grid search or random search \cite{Bergstra:2012:RSH:2188385.2188395}
over the hyperparameter space.

\subsection{Technologies Deployed}

All the experiments were carried out on a server housing 48 core Intel
Xeon @ $2.5$ GHz processor, 252 GB RAM with Ubuntu 16.04. The experiments
were coded in Python 3.6, using 3rd party libraries - NetworkX \cite{hagberg-2008-exploring},
Numpy \cite{Walt:2011:NAS:1957373.1957466}, Matplotlib \cite{Hunter:2007},
Scikit-learn \cite{scikit-learn} and Gensim \cite{rehurek_lrec}. 

\section{Conclusion\label{sec:Conclusion}}

We have presented a novel supervised Q-walk approach to generate random
walks guided by Q-values and assisted by k-hops neighborhood based
confidence values learner. We have shown experimentally that the node
embeddings learnt from our approach are similar for the nodes with
the same labels. We have also shown that our approach outperforms
node2vec in the node classification task. 

\section{Future Work\label{sec:Future-Work}}

Supervised Q-walk works better in the cases where the assumption of
homophily in networks holds true. In other networks, nodes which are
structurally equivalent may have the same labels e.g. networks of
bio-chemical compounds. Our work can be extended by composing another
reward function $R(u,a,u')=R_{1}(u,a,u')+\beta R_{2}(u,a,u')$ where
$R_{1}$ encourages homophily in networks, $R_{2}$ encourages structural
equivalence of nodes and $\beta$ is a hyperparameter which decides
the tradeoff between homophily and structural equivalence. 

\section{Acknowledgements}

We acknowledge the valuable insights provided by Dr Robert West, Assistant
Professor, Data Science Lab, School of Computer and Communication
Sciences, EPFL, Switzerland. His lab had also provided us with the
compute infrastructure for carrying out all the experiments.

\bibliographystyle{plain}
\bibliography{supervised_q_walk}

\end{document}